\documentclass[preprint,showpacs,preprintnumbers]{revtex4}%

\usepackage{amsfonts}
\usepackage{amsmath}
\usepackage{amssymb}
\usepackage{graphicx}

\begin{document}

\title{Thermospin Hall effect generated by thermal influence and
thermoelectric effect}
\author{Zhongshui Ma}
\affiliation{School of Physics, Peking University, Beijing 100871,
China}

\begin{abstract}
In this paper, we present the theoretical predication of a thermospin Hall
effect, in which a transverse spin current can be generated in
semiconductors in the presence of spin-orbit coupling by a
frequency-dependent longitudinal temperature gradient. The thermospin Hall
effect has a number of qualitative distinctions in comparison with the spin
Hall effect driven by an electric field. Because of the thermoelectric
effect, there is no net charge current but there is a heat flow from the hot
side to the cold side. We perform the theoretical calculation of dynamical
thermospin Hall conductivity in a two-dimensional Rashba spintronic system.
It has been shown that the direct interband optical transition dominates the
ordering and manipulation of spin in the generation of a transverse
intrinsic spin current. In view of the role of the thermoelectric effect,
the contributions to the thermospin Hall effect are classified as that
originating from a direct contribution of thermal electronic diffusion and
that from the compensatory electron flow in balance with the thermal
diffusion. In physical terms, we explain the phenomenon as the spin-orbit
coupling exerting force on electronic orbital motions, which are driven by
the thermoelectric properties, and manipulating the
spin-orientation-dependent motions. For a finite system, the analysis yields
evidence that the spin accumulation around the edges of a plate determines
the magnetization. In equilibrium, a field created by a magnetization
gradient emerges in the direction perpendicular to the temperature gradient.
The experimental observation of the thermospin Hall effect is proposed by
measuring the longitudinal temperature difference with the injection of a
transverse spin current and by analyzing the Hall angle. In addition, in
order to achieve pure spin accumulation in the spin Hall effect, an
extension of the thermospin Hall effect for exciting electron-hole pairs in
semiconductors is proposed.
\end{abstract}

\maketitle

Rapid developments in technology and manufacturing are providing
us with favorable circumstances for fabricating high-speed and
low-powered electronic devices that use the spin properties of
electrons. This will alter the current situation that microchips
use only the charge properties of electrons. In order to
efficiently utilize the advantageous functions of spin-based
recording and information processing, one of the crucial issues is
to accomplish pure spin injection virtually$^1$. In fact, besides
its potential applications in spintronic devices$^2$, the
electrical generation and manipulation of spin flux in
semiconductors also has its own fundamental physics worthy of
being studied. Among such studies, the spin Hall effect in
narrow-band semiconductors has been a focal point of research in
last few years$^{3-6}$. As a quantum mechanical degree of freedom
attached to electrons, the spin can couple with the orbital motion
of an electron through an internal electric field. A simple image
of this coupling may be obtained as follows: The orbital motion of
an electron in an internal electric field creates a magnetic field
in its vicinity. On account of the electron spin, an electron has
an intrinsic magnetic moment. The interaction between the electron
spin magnetic moment and the magnetic field created by its orbital
motion is regarded as the spin-orbit (SO) coupling. The SO
interaction energy leads to a spin-splitting in the energy
spectrum of moving electrons, even in the absence of any magnetic
field. For a two-dimensional electron gas (2DEG) in a
heterostructure, such as InAs- and In$_{1-x}$Ga$_{x}$As quantum
wells, the structural inversion asymmetry of the confining
potential due to the presence of the heterojunction results in the
spin splitting of the conduction band in momentum (k) space$^7$.
The formation of spin-splitting bands due to Rashba SO
coupling$^8$ has been measured in a number of materials, e.g., in
heterostructures based on InAs$^9$ or HgTe$^{10}$. Experimentally,
Rashba SO coupling can be changed externally, e.g., by applying
additional back-gate voltage to the structure$^{11}$. The change
in the strength of SO coupling induces a modulation of electronic
band structures, which is equivalent to the manipulation of
electron spins. Hence, the spin-orientation-dependent motion can
eventually be controlled by regulating the voltage gate without a
magnetic field. Similar to a magnetic field exerting Lorentz force
transversely on the moving charge, the SO coupling produces a
transverse "force" on the moving spin$^{12-14}$. Under the spin
"force," electrons with opposite spin orientations drift off their
initial direction of motion and tend to separate spatially in
opposite directions. If the motion of electrons is driven by an
external electric field, a result is that a spin current is
generated in the direction perpendicular to the electric field,
without an accompanying transverse charge current. Differing from
the phenomenon that an extrinsic mechanism causes the spatial
separation of electrons with different spin orientations, the
phenomenon that is merely due to the SO coupling in solids is
subsequently referred to as the intrinsic spin Hall
effect$^{5,6}$. Recently, related studies have been extended to
address the quantum spin Hall effect by employing the existence of
bulk gap and gapless edge states in a time-reversal invariant
system with SO coupling$^{15,16}$.

It is common knowledge that an electric current can be generated
not only by an applied electric field but also by a temperature
gradient in solids$^{17}$. This gives rise to a great number of
interesting thermoelectric phenomena. A particularly interesting
transport property is the thermoelectric (Seebeck) effect, i.e.,
the temperature gradient produces an electric potential gradient
which can drive an electric current. Due to the extreme
sensitivity of 2DEG to changes in electronic structure at the
Fermi energy$^{18}$, the thermoelectric powers of 2DEG in
AlxGa1-xAs/GaAs heterojunctions at low temperatures have attracted
much attention$^{19}$. Studies show that the thermal features of
semiconductor materials can play an important role in transport.
In practice, a sensitive probe of the transport mechanism has
included some proposals based on the use of thermoelectric power.
One can, therefore, expect that this mechanism should play a role
in semiconductor spintronic materials and will provide a
challenging opportunity in developing new thermospintronic
devices, whose thermoelectric properties can be controlled by SO
coupling.

Here we introduce a new "member" of the spin Hall effect "family" - the
thermospin Hall effect. The present study shows that a transverse spin
current can be generated in a 2DEG in the presence of SO coupling by a
longitudinal thermal gradient. Correspondingly, thermospin Hall conductivity
is defined as the ratio between the transverse spin current and the
longitudinal thermal gradient. We schematically illustrate these phenomena
in Figure 1(a). A heuristic picture of the thermospin Hall effect is the
combination of the thermal diffusion of electrons, the thermoelectric
effect, and the spin-orientation-dependent side drift driven by SO coupling.
When there exists the time-dependent temperature gradient $\mathbf{\nabla }T$
(parallel to the $\widehat{\mathbf{x}}$ direction) in 2DEG, an electric
potential distribution is built up by the Seebeck effect. The spatial
distribution of the electric potential induces a backflow of electrons,
which tends to balance the charge current driven by $\nabla _{x}T$.
Meanwhile, the time-dependent electric potential stimulates electronic
transitions between spin-splitting bands. As a result, no net average
electric current flows in the $\widehat{\mathbf{x}}$ direction. Since moving
electrons have not only charge but also spin, the SO coupling in the system
exerts a spin force on moving spins in the $\widehat{\mathbf{y}}$ direction,
i.e., perpendicular to the temperature gradient. The force depends on both
the spin orientation and the direction of electron orbital motion. Electrons
with opposite spins are forced to move oppositely in the $\widehat{\mathbf{y}%
}$ direction. Hence, while not accompanied by net charge current, a pure
spin current, associated with electronic transitions between spin-splitting
bands, is generated. To differentiate from the spin Hall effect driven by an
electric field, we regard this phenomenon as the thermospin Hall effect.
Here the thermoelectric effect and the transition between spin-splitting
states play important roles. Regarding this phenomenologically, the spin
current can, in general, be expressed in the form $\mathbf{J}^{\left(
s\right) }=\mathbb{L}^{\left( se\right) }\cdot \mathbf{E}+\mathbb{L}^{\left(
sq\right) }\cdot \left( -\mathbf{\nabla }T/T\right) $, where $\mathbb{L}%
^{\left( se\right) }$ and $\mathbb{L}^{\left( sq\right) }$ are the spin
conductivity tensor and the thermospin conductivity tensor, respectively.
The induced electric field is determined by the balancing of the
longitudinal electric current, i.e., $E_{x}=S\nabla _{x}T$ with Seebeck
coefficient $S$. In the linear response, the corresponding spin current in
the $\widehat{\mathbf{y}}$ direction can be written as\textbf{\ }$%
J_{y}^{z\left( s\right) }=\sigma _{TH}^{SH}\nabla _{x}T$, where
$\sigma _{TH}^{SH}=\left( TSL_{yx}^{\left( se\right)
}-L_{yx}^{\left( sq\right) }\right) /T$ is the thermospin Hall
conductivity.

To formulate thermospin Hall conductivity, it is necessary to
calculate the nonequilibrium carrier current density influenced by
electric potential and temperature gradients. The crucial point in
studying the spin thermal transport is to clarify the influence of
temperature in the microscopic theoretical description. There is a
long history of microscopic study in thermoelectric transport
phenomena. Since temperature is a statistical property of the
system, there is no Hamiltonian to describe the thermal gradient.
Therefore, although the linear response to an external electric
field leads unambiguously to the Kubo formula for the electric
spin Hall conductivity tensor, its extension to the calculation of
thermospin Hall conductivity is not straightforward. Fortunately,
in 1964, Luttinger provided microscopic proof that thermoelectric
transport coefficients are given by the corresponding
current-current correlation function$^{20}$. In the framework of a
so-called "mechanical" derivation, an inhomogeneous gravitational
field is introduced to produce energy flow and temperature
fluctuation$^{19-21}$. Theoretically, an energy density $H\left( \mathbf{r}%
\right) $ behaves as if it had a mass density $H\left(
\mathbf{r}\right) /c^{2}$, which interacts with the gravitational
field $-\left( 1/c^{2}\right) \psi \left( \mathbf{r},t\right) $.
Varying $\psi $ will cause an energy current to flow. Further, a
varying energy density gives rise to a temperature gradient. The
macroscopic currents arising in a nonequilibrium system are
proportional to the driving forces $\mathbf{E}-\left( T/e\right)
\mathbf{\nabla }\left( \mu /T\right) $ and $T\mathbf{\nabla
}\left( 1/T\right) -\mathbf{\nabla }\psi $, where $-\left(
T/e\right) \mathbf{\nabla }\left( \mu /T\right) $ ($\mu $ is
chemical potential and $e$ is electron charge) and
$T\mathbf{\nabla }\left( 1/T\right) $ are statistical forces,
while $\mathbf{E}$ and $-\left( 1/c^{2}\right) \mathbf{\nabla
}\psi $ are the electric field and gradient of the gravitational
field, respectively. Due to the induced temperature gradient, a
compensating energy current flowing in the opposite direction
brings the system into equilibrium. In this way, the variation of
the added energy fluctuation is balanced by a
temperature distribution. This leads to the identification $\mathbf{\nabla }%
\psi =\mathbf{\nabla }T/T$. Einstein's relationship is tenable - it relates
the response to electrical field $\mathbf{E}$ and gravitational field
gradient $\mathbf{\nabla }\psi $ to the observed concentration gradient $%
\left( T/e\right) \mathbf{\nabla }\left( \mu /T\right) $ and temperature
gradient $T\mathbf{\nabla }\left( 1/T\right) $, respectively. In
equilibrium, zero current conditions lead to the relations $\mathbf{E}%
=\left( T/e\right) \mathbf{\nabla }\left( \mu /T\right) $ and $T\mathbf{%
\nabla }\left( 1/T\right) =\mathbf{\nabla }\psi $. The transport
coefficients response to $T\mathbf{\nabla }\left( 1/T\right) $ equals that
to $\mathbf{\nabla }\psi $ and the same is true for the coefficients of $%
\mathbf{E}$ and $\left( T/e\right) \mathbf{\nabla }\left( \mu /T\right) $.
Hence, it is only necessary to consider the system response to dynamical
forces for calculations of transport coefficients. This theoretical
description is then valid for spin-dependent transport. The linear response
theory enables us in practice to perform analytic derivations of spin Hall
conductivity in the case of non-uniform temperature $T$. In the presence of
SO coupling, the spin-orientation dependence has been found to be involved
in the interaction between energy density and the gravitational field.
Generalizing Einstein relations to the spin-dependent response theory, the
spin-dependent thermal coefficients can be obtained on the analogy of
spin-dependent (charge and spin) conductivities.

We now present a theoretical calculation of the spin Hall current
generated by thermal influence and the thermoelectric effect. Here
we consider a two-dimensional Rashba spintronic system in the
presence of a time-dependent temperature gradient$^{21-23}$, i.e.,
studying the dynamical response to a temperature gradient
alternating with nonzero frequency. This yields
frequency-dependent thermopower and dynamical thermospin Hall
conductivity. The Hamiltonian in the presence of infinitesimal
time-dependent electric
field $\mathbf{E}(\mathbf{r},t)=-\mathbf{\nabla }\varphi \left( \mathbf{r}%
,t\right) $ and gradient of the gravitational field $\nabla \psi \left(
\mathbf{r},t\right) $ is given by $H=H_{0}+\left( 1/c\right) \mathbf{J}%
^{\left( e\right) }\cdot \mathbf{A}\left( \mathbf{r},t\right) +\left(
1/e\right) \mathbf{J}^{\left( q\right) }\cdot \mathbf{N}\left( \mathbf{r}%
,t\right) $, where vector potentials $\mathbf{A}(\mathbf{r},t)=\left(
c/i\omega \right) \mathbf{E}(\mathbf{r})e^{-i\omega t+0^{+}t}$ and $\mathbf{N%
}\left( \mathbf{r},t\right) =\left( e/i\omega \right) \nabla \psi \left(
\mathbf{r}\right) e^{-i\omega t+0^{+}t}$, which are adiabatically switched
on from the infinitely remote past $t=-\infty $, interact with electrical
and energy currents, respectively. $H_{0}$ is the Rashba Hamiltonian and can
be written in a second quantized form, $\sum_{\mathbf{k},s}E_{\mathbf{k}%
,s}^{(0)}a_{\mathbf{k},s}^{\dag }a_{\mathbf{k},s}$, where $a_{\mathbf{k},s}$
($a_{\mathbf{k},s}^{\dag }$) is the annihilation (creation) operator for an
electron with momentum $\hbar \mathbf{k}$ and band $s$. $s=\pm $ denotes two
spin-splitting dispersion branches' energy $E_{\mathbf{k},\alpha
}^{(0)}=\hbar ^{2}k^{2}/2m^{\ast }+s\lambda k$ with Rashba SO coefficient $%
\lambda $, the effective electron mass $m^{\ast }$ and the in-plane momentum
$\hbar k=\hbar \sqrt{k_{x}^{2}+k_{y}^{2}}$. Corresponding electric current
operator $\mathbf{J}^{\left( e\right) }$ and heat current operator $\mathbf{J%
}^{\left( q\right) }$ take the form
\begin{equation*}
\mathbf{J}^{\left( e\right) }=e\sum_{\mathbf{k},s}\mathbf{K}_{\mathbf{k}%
,\parallel }^{\left( s\right) }a_{\mathbf{k},s}^{\dag }a_{\mathbf{k}%
,s}-ie\sum_{\mathbf{k},s}\mathbf{K}_{\mathbf{k},\perp }^{\left( s\right) }a_{%
\mathbf{k},s}^{\dag }a_{\mathbf{k},-s}
\end{equation*}%
and%
\begin{equation*}
\mathbf{J}^{\left( q\right) }=\sum_{\mathbf{k},s}\left( E_{\mathbf{k}%
,s}^{(0)}-\mu \right) \mathbf{K}_{\mathbf{k},\parallel }^{\left( s\right)
}a_{\mathbf{k},s}^{\dag }a_{\mathbf{k},s}-i\sum_{\mathbf{k},s}\left( \frac{%
\hbar ^{2}k^{2}}{2m^{\ast }}-\mu \right) \mathbf{K}_{\mathbf{k},\perp
}^{\left( s\right) }a_{\mathbf{k},s}^{\dag }a_{\mathbf{k},-s}
\end{equation*}%
with $\mathbf{K}_{\mathbf{k},\parallel }^{\left( s\right) }=\left( \hbar
/m^{\ast }\right) \mathbf{k}\left[ 1+s\left( \lambda m^{\ast }/\hbar
^{2}k\right) \right] $ and $\mathbf{K}_{\perp }^{\left( s\right) }=s\left(
\lambda /\hbar \right) \left[ \left( \mathbf{k}\times \widehat{\mathbf{z}}%
\right) /k\right] $. From the expressions of currents it can be
seen that, responding to the frequency-dependent electric field
and temperature gradient, the contributions to currents are
composed of two parts: the intraband current and the interband
current. The latter comes mainly from the anomalous velocity term
related to the transition between spin-splitting subbands. To
pinpoint the specific details of subband transitions that are
responsible for the generation of the spin Hall current, we keep
the discussion within "clean" limits, specifically that the system
has no disorder, impurity, or electron-phonon
interaction$^{24,25}$. In evaluating spin states with
spin-splitting subbands, the equation of motion in the
linear response to $\mathbf{A}$ and $\mathbf{N}$ leads to $\left\langle a_{%
\mathbf{k},s}^{\dag }a_{\mathbf{k},s}\right\rangle ^{\left( 1\right) }=0$ and%
\begin{equation*}
\left\langle a_{\mathbf{k},s}^{\dag }a_{\mathbf{k},-s}\right\rangle ^{\left(
1\right) }=-is\frac{\lambda }{\hbar k}F_{\mathbf{k}}^{s,-s}\left( \omega
\right) \left[ \mathbf{k}\times \left( \frac{e}{c}\mathbf{A+}\frac{\hbar
^{2}k^{2}}{2m^{\ast }e}\mathbf{N}\right) \right] \cdot \widehat{\mathbf{z}},
\end{equation*}%
where $F_{\mathbf{k}}^{s,-s}\left( \omega \right) =\left( f_{\mathbf{k}%
,s}-f_{\mathbf{k},-s}\right) /\left( \hbar \omega ^{\prime }+E_{\mathbf{k}%
,s}^{(0)}-E_{\mathbf{k},-s}^{(0)}\right) $ with $\omega ^{\prime }=\omega
+i0_{+}$ and Dirac-Fermi distribution function $f_{\mathbf{k},s}$. These
mean that only the interband transitions cause dynamical currents while
intraband terms do not contribute to conduction if the system is free of any
disorders. After a straightforward calculation, the frequency-dependent
currents in the linear response are found as $\mathbf{J}^{\left( n\right)
}\left( \omega \right) =\mathbb{L}^{\left( ne\right) }\left( \omega \right)
\cdot \mathbf{E}+\mathbb{L}^{\left( nq\right) }\left( \omega \right) \cdot
\left( -\mathbf{\nabla }T/T\right) $, where $n=1$ denotes charge ($e$) and $%
n=3$ denotes heat ($q$). The corresponding transport coefficients, the order
of $\lambda ^{2}$, can be cast in the from $\mathbb{L}\left( \omega \right)
\propto \sum_{\mathbf{k},s}\mathbb{G}_{\mathbf{k},s}^{\left( n\right)
}\left( \omega \right) $ with $\mathbb{G}_{\mathbf{k},s}^{\left( n\right)
}\left( \omega \right) =k^{n}F_{\mathbf{k}}^{s,-s}\left( \omega \right)
\mathbb{I}$ and two dimensional unit tensor $\mathbb{I}=\widehat{\mathbf{x}}%
\widehat{\mathbf{x}}+\widehat{\mathbf{y}}\widehat{\mathbf{y}}$. For $%
E_{y}=\nabla _{y}T=0$ and $\nabla _{x}T\neq 0$, $\mathbf{j}^{\left( e\right)
}=0$ is a result of the complete cancellation in opposite electric currents
owing to electrons flowing between hot and cold sides in the system. As a
consequence of the thermoelectric effect, a longitudinal electric field $%
E_{x}$ is induced by the temperature gradient, i.e., $E_{x}=S\nabla _{x}T$,
where $S=L^{\left( eq\right) }/\left( TL^{\left( ee\right) }\right) $ is the
Seebeck coefficient.

Accompanying the electron motion in a Rashba SO coupling system, a spin
force, oriented perpendicular to the temperature gradient, acts on its spin.
Although the net electric current vanishes, the number of electrons moving
in opposite directions is not of an equal amount. The excess of moving
electrons will contribute to generating the spin Hall current. The drift
processes of electrons are opposite with respect to their spin orientations
being up or down. In that the opposite spins move in opposite directions,
this sets up a transverse spin current in the system. The spin current
operator in the second quantized form is given by $\mathbf{J}^{z\left(
s\right) }=\left( \hbar ^{2}/2m^{\ast }\right) \sum_{\mathbf{k},s}\mathbf{k}%
a_{\mathbf{k,}s}^{\dag }a_{\mathbf{k,-}s}$. It is quite evident that the
spin current is generated not by the displacement from the electron
distribution function, but by the contribution from transitions between
spin-split bands. The calculation shows that the only nonzero component of
the spin current is in the $\widehat{\mathbf{y}}$ direction, i.e.,
perpendicular to the temperature gradient. It is noted that the
contributions to the spin current are naturally classified into a direct
thermal spin Hall current, caused by a temperature gradient $-\left(
L^{\left( sq\right) }/T\right) \nabla _{x}T$, and a thermoelectric spin Hall
current, by thermoelectric effect $SL^{\left( se\right) }\nabla _{x}T$,
where $L^{\left( sq\right) }$ and $L^{\left( se\right) }$ relate to $s%
\mathbb{G}_{\mathbf{k},s}^{\left( 2\right) }\left( \omega \right) $ and $s%
\mathbb{G}_{\mathbf{k},s}^{\left( 4\right) }\left( \omega \right) $,
respectively, and are in the order of $\lambda $. The total spin Hall
current can be written, in terms of the thermospin Hall conductivity $\sigma
_{TE}^{SH}=\left( TSL^{\left( se\right) }-L^{\left( sq\right) }\right) /T$,
as $J_{y}^{z\left( s\right) }=\sigma _{TE}^{SH}\nabla _{x}T$.

Figure 2 shows the temperature dependence of thermospin Hall conductivity
against the strength of SO coupling. Here a finite frequency is chosen and
the thermospin Hall conductivity is frequency dependent. It is readily seen
that the thermospin Hall conductivity $\sigma _{TH}^{SH}\left( \omega
\right) $ is non-vanished when the frequency $\omega $ is in the region $%
2\lambda k_{F,+}\leq \hbar \omega \leq 2\lambda k_{F,-}$, where $k_{s}=\sqrt{%
2m^{\ast }\lambda ^{2}/\hbar ^{2}}\sqrt{4\mu +2m^{\ast }\lambda ^{2}/\hbar
^{2}}-s2m^{\ast }\lambda ^{2}/\hbar ^{2}$. The interband transitions mainly
appear between energies $\hbar \omega _{+}=2\lambda k_{F,+}$ and $\hbar
\omega _{-}=2\lambda k_{F,-}$, which correspond respectively to the minimum
and maximum photon energy required to induce optical transitions between the
initial $s=-1$ and final $s=+1$ spin-split bands. The width of the resonant
window is almost independent of temperature because the change of
temperature is merely broadening the range of electron distribution at the
Fermi energy. However, increasing temperature shifts the resonant window
toward the high-frequency regime. In the inset of Figure 2, we show the
shifting of the resonant frequency window with respect to altering the
strength of the SO coupling. As seen in Figure 2, the magnitude of
thermospin Hall conductivity depends on the temperature and displays as
hyperbolic in nature. It exhibits the behavior $1/T$ at low temperature if
the frequency is within the range ($\omega _{+}$, $\omega _{-}$). When the
temperature increases, the thermospin Hall conductivity decreases rapidly,
while at high temperature, it tends slowly to a frequency-dependent low
finite value. In contrast, if the frequency is outside the range ($\omega
_{+}$, $\omega _{-}$), the thermospin Hall conductivity tends to zero at low
temperature. With increasing temperature, the thermospin Hall conductivity
increases first and achieves a maximum value at a certain temperature. With
further increasing temperature, the thermospin Hall conductivity will
decrease and tend to a frequency-dependent low finite value. These findings
indicate that the measurement of thermospin Hall current can be controlled
by modulating electrical gating and by adjusting the frequency of the
temperature gradient. The observation of a relatively strong magnitude in
the frequency domain hinges on the appropriate temperature.

If the 2DEG is confined as a plate with a constant gate voltage applied in
the $\widehat{\mathbf{z}}$ direction (controlled strength of SO coupling),
as illustrated in Figure 1(a), we attach its two ends to heat baths with
high temperature $T_{H}$ on the left and low temperature $T_{L}$ on the
right (in the $\widehat{\mathbf{x}}$ direction). The heat current,
associated with moving electrons, flows between the left and the right ends.
The SO coupling manifests as a spin force on the spin of electrons in the
direction perpendicular to the heat flow. As a result, the formation of
spin-orientation-dependent electronic motion leads to spins with opposite
signs preferentially deviating in opposite directions. An excess of up spins
will be accumulated on one side of the plate and an excess of down spins on
the opposite side. The spin alignment of electrons (ferromagnetic profiles)
on the edge determines the magnetization, i.e., the magnetization $%
M_{\uparrow }$ is distributed in the region near one edge and $M_{\downarrow
}$ near another edge. The magnetizations in the $+\widehat{\mathbf{z}}$ and $%
-\widehat{\mathbf{z}}$ directions produce stray fields in the $+\widehat{%
\mathbf{y}}$ and $-\widehat{\mathbf{y}}$ directions, respectively. The field
exerts the force on the incoming spin-polarized flow through the interaction
$\mu _{B}j_{y}^{z}\nabla _{y}M_{z}$ and stops any further spin exchange.
Equilibrium is created due to the balancing of the spin force of SO
coupling, the electric force due to charge accumulation, and the force owing
to the gradient of magnetization. The spin accumulation profiles persist
after the spin currents have vanished. Then, in the $\widehat{\mathbf{y}}$
direction, a potential distribution $V_{SH}$ is generated on the upper
half-plane and the voltage $-V_{SH}$ on the lower half-plane, as illustrated
in Figure 1(a). SO coupling is equivalent to the effective SU(2) gauge
potential, transverse to the temperature gradient in the Pauli Hamiltonian.
As such, similar to the Nernst effect in a conductor plate, a potential
distribution built up from stray magnetization can emerge in the $\widehat{%
\mathbf{y}}$ direction under SO coupling (instead of a magnetic field in the
$\widehat{\mathbf{z}}$ direction in the Nernst effect) and a temperature
bias in the $\widehat{\mathbf{x}}$ direction.

The predication of an effective potential caused by a
magnetization gradient opens up possibilities for measuring the
thermospin Hall signal. One such possibility is that, to confirm
the thermospin Hall effect experimentally, the characteristics of
light reflected in magnetizations owing to spin alignment can be
sought. Recently, several groups reported optical detections of
spin accumulation with opposite signs at the sample edges in
current-biased nonmagnetic semiconductors$^{26-28}$. In these
experiments, Rashba-type spin splitting can be identified via
optical experiments instead of magnetotransport measurements.
Optical experiments have also been used to study the generation of
a pure spin current$^{29,30}$. Another possibility that emerges is
the spin accumulation induced by a thermoelectric effect,
providing us a reciprocal way to confirm the thermospin Hall
effect. In fact, the predication of the thermospin Hall effect
offers us a new pathway to obtain spin information by means of the
detection of longitudinal temperature differences when a spin
current is transversely injected. As a transverse spin current is
injected in a nonmagnetic semiconductor, the effect is essentially
that electrons with opposite spins flow in opposing transverse
directions. The SO interaction in the system leads electrons to
move toward the same longitudinal side of the sample. This results
in a charge accumulation and induces a longitudinal temperature
gradient by means of the thermoelectric effect. In fact, similar
electrical measurements, based on the idea of the inverse spin
Hall effect, have been achieved with a charge accumulation$^{31}$.
In addition, due to the opening up of new channels for thermospin
transition accompanied by heat-radiating absorption and
excitation, some unique thermal-optical properties can be observed
in thermospin Hall systems$^{32}$. Different from experiments
where the induced magnetic moment is non-destructively detected in
a non-contacting way using a magnetometer$^{33}$, the thermospin
Hall effect provides a promising method for detecting
inhomogeneity of spin accumulation in semiconductor materials,
employing a thermistor probe and a thermogalvanometer.

It becomes of interest to us to decide whether the thermal effect in
heterogeneous semiconductors might be of use in practical spintronics.
Spin-splitting states are asymmetric in the momentum distribution of
electrons in Rashba spintronic systems and can usually be controlled
experimentally by tuning external parameters. Therefore, the topic of the
thermospin properties of Rashba spintronic systems, such as the
thermoelectric-driven spin current proposed in this study, is very rich in
terms of basic physics and device applications. Studying the spin-dependent
Hall angle allows us to address the problem of the validity of
spin-orientation-dependent edge states in the spin Hall effect. The Hall
angle $\Theta $ is defined by $\tan \Theta =E_{y}/E_{x}$, where $E_{y}$ and $%
E_{x}$ are the transverse and longitudinal potential gradients,
corresponding respectively to completely different mechanisms: the former
stems from the induced spin accumulation due to the SO coupling, and the
latter, in essence, originates from the thermoelectric effect. The Hall
angle in equilibrium is a reflection of the thermospin and thermoelectric
effects of a given system. Because the spin experiences a spin force ($%
\propto \mathbf{j}^{z\left( s\right) }\times \widehat{\mathbf{z}}$), we note
that the spin diffusion process is rotated away from the $\widehat{\mathbf{x}%
}$ direction by a Hall angle $\Theta $. Under the interaction between the
spin-polarized flow and the gradient field of magnetization, a backflow
current $J_{x}^{\left( e\right) }=\mu _{B}L^{\left( es\right) }\nabla
_{y}M_{z}$ is generated with the aid of an anomalous Hall/converse spin Hall
effect. It is convenient to regard this as an anomalous electric Hall effect
induced by the magnetization gradient. This backflow electric current is
rotated away from the $-\widehat{\mathbf{x}}$ direction by the Hall angle $%
\Theta $. It is worthwhile to point out that the generation of a
Hall-like backflow electric current originates entirely from the
interaction $\mu _{B}j_{y}^{z}\nabla _{y}M_{z}$ breaking the
time-reversal symmetry$^{34}$. In equilibrium, $J_{x}^{\left(
e\right) }=J_{y}^{z\left( s\right) }=0$ yields for the
longitudinal $\left( E_{x}\right) $ and transverse $\left(
E_{y}\right) $ potential gradients, $E_{x}=\left( L^{\left(
sq\right) }/TL^{\left( se\right) }\right) \nabla _{x}T$ and
$E_{y}$ $=-\left( \hbar /2e\right) \left[ \left( L^{\left(
eq\right) }L^{\left( se\right) }-L^{\left( ee\right) }L^{\left(
sq\right) }\right) /TL^{\left( se\right) }L^{\left( se\right)
}\right] \nabla _{x}T$, respectively. Therefore, we obtain $\tan
\Theta =\left( \hbar /2e\right) \left[ \left( L^{\left( ee\right)
}L^{\left( sq\right) }-L^{\left( eq\right) }L^{\left( se\right)
}\right) /L^{\left( se\right) }L^{\left( sq\right) }\right] $.
Figure 3 presents an estimation of the Hall angle in the
thermospin Hall effect. The most prominent features of frequency
dependence of the Hall angle are sharp divergences around $\hbar
\omega _{\pm }=2\lambda k_{F,\pm }$ (Figure 3(b)). The absolute
value decreases rapidly as $\omega $ deviates from $2\lambda
k_{F,\pm }$. The distance between divergences is temperature
dependent (Figure 3(a)). The reason is that both spin transport
and the magnetization on the edges are dominated by those
electrons which are thermally created in relatively wider energy
bands. The sign of the Hall angle depends on whether the frequency
is inside or outside the frequency range $\Delta \omega =(\omega
_{-},\omega _{+})$. In addition, the transverse potential gradient
decreases as temperature increases. The reason is that, in
principle, the spin alignment decreases as the contribution of
spin splitting is suppressed by thermal energy with the increase
of temperature. In practice, the spin diffusion length also slowly
decreases at high temperature.

Finally, we should point out that a charge accumulation
accompanies the electronic spin Hall effect, due to an excess of
electrons with the same spin orientation gathering around edges.
Such a charge population might produce an influence in optical
measurements. To eliminate the influence caused by the charge
accumulation, it is essential to achieve pure spin accumulation in
the spin Hall effect. This can be realized by the
thermophotovoltaic generation of electron-hole pairs in
semiconductor systems. This is similar to the optical schemes for
generating spin current$^{29,30}$, which are also valid for the
thermospin Hall effect. For situations in which the Fermi level
lies between the conduction and valence bands, the absorption of
"photon" energy excites an electron in the valence band to move
into the conduction band. As a result, an electron-hole pair is
created. The charge currents due to electrons and holes under the
temperature gradient are opposite and equal in magnitude. In this
case, there is no net electric current. Correspondingly, the
thermopower is negative for electrons and positive for
holes$^{35}$. As it has been described for Rashba systems, the
spins of both moving electrons and holes exert spin force. In
identifying the pair formed by an electron and hole with opposite
spins, it can be seen that the forces exerted on them are
opposite. Such spin asymmetry of an electron-hole pair results in
the pair breaking and the electron and hole separating spatially
under the spin force. In narrow-gap insulators, the electron-hole
interaction is very weak, due to sufficient screening. In these
systems, excitonic effects do not dominate the spectrum and
one-electron spectrum approximation is applicable. Therefore,
electrons and holes with the same spin orientation will deviate in
the same direction. As illustrated in Figure 1(b), the motions of
electrons and holes from the central excitation spot upward for up
spins and downward for down spins results in the spin current. In
this way, the opposing charge currents due to electrons and holes
are converted into the intrinsic spin current. There also exist
two kinds of contributions in the thermospin Hall effect: one is
the direct contribution from spin asymmetry in the electron-hole pair $%
-\left( L^{\left( sq\right) }/T\right) $ and another is from the
thermoelectric effect $SL^{\left( se\right) }$. For a plate, there
is only net magnetization but no net charge in edges because the
populations of electrons and holes are equal and have the same
spin orientations. No charge distribution in the
$\widehat{\mathbf{y}}$ direction means no electric field is
generated in the $\widehat{\mathbf{y}}$ direction either. Hence,
no Joule heating is caused. This characteristic is particularly
suitable, not only for a dissipationless experimental measurement,
but also for device applications at low power loss. We emphasize
that the survival of thermospin Hall phenomena in semiconductors
strongly depends upon the spin-splitting band structure, which
will be influenced in principle by the phonon spectrum$^{36}$ and
the peculiarities of the scattering mechanisms$^{37}$. The
electron-phonon interaction in the thermospin Hall effect are
interesting in practice because the thermoelectric effect would be
drastically changed by the electron-phonon interaction, ranging
from the weak-coupling limit$^{38}$ to the strong-coupling
regime$^{39}$. This may give further insight both for the nature
of the thermospin Hall effect and for spintronic applications.

\begin{acknowledgments}
We thank L. DesBrisay, T.H. Lin, J.F. Liu, G.S. Tian, S.H. Zhu,
and J. Wang for valuable discussions. This work is supported by
NNSFC Grant No. 10674004 and NBRP-China Grant No. 2006CB921803.
\end{acknowledgments}

\textbf{Figures}

\begin{figure}[tbp]
\centering \includegraphics[width=12.0cm]{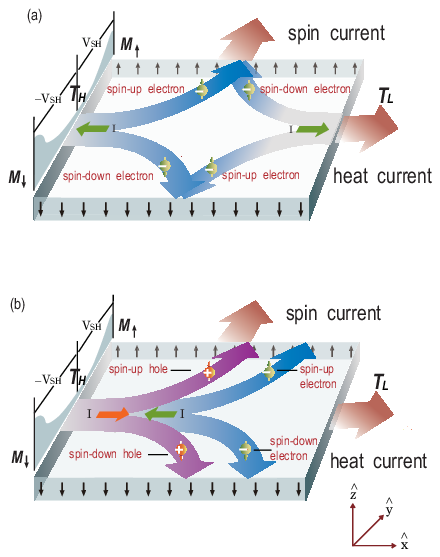}
\caption{Schematic illustration of a traverse
spin-orientation-dependent motion in SO coupling systems generated
by longitudinal temperature influence: (a) electron gas and (b)
the scheme proposed for excitation of electron-hole pairs. The
left and right ends of two-dimensional systems are contacted to
the periodically modulated hot and cold baths, respectively. The
heat current, associated with moving electrons, flows between the
left and right ends. The corresponding charge current is balanced
by the thermoelectric effect. The spin-up (spin-down) carriers are
shown moving in the $+\widehat{\mathbf{y}}$
($-\widehat{\mathbf{y}}$) direction, yielding no transverse net
charge current. As a result, a spin current, without accompanying
net charge current, is generated in the direction perpendicular to
the temperature gradient. The magnetizations in opposite out-of
plane orientations are formed in the opposite edges if the systems
are plates.} \label{fig1}
\end{figure}

\begin{figure}[tbp]
\centering \includegraphics{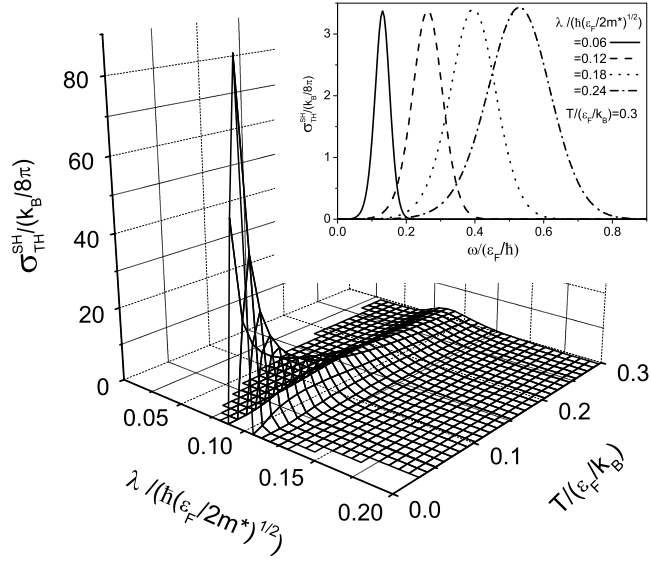} \caption{The thermospin Hall
conductivity versus the strength of SO coupling and temperature in
the presence of a periodically modulated temperature gradient.
Here the frequency has been taken 0.2 in the unit of Fermi energy.
The inset shows the frequency-dependent thermospin Hall
conductivity with five different strengths of SO coupling.}
\label{fig2}
\end{figure}

\begin{figure}[tbp]
\centering \includegraphics{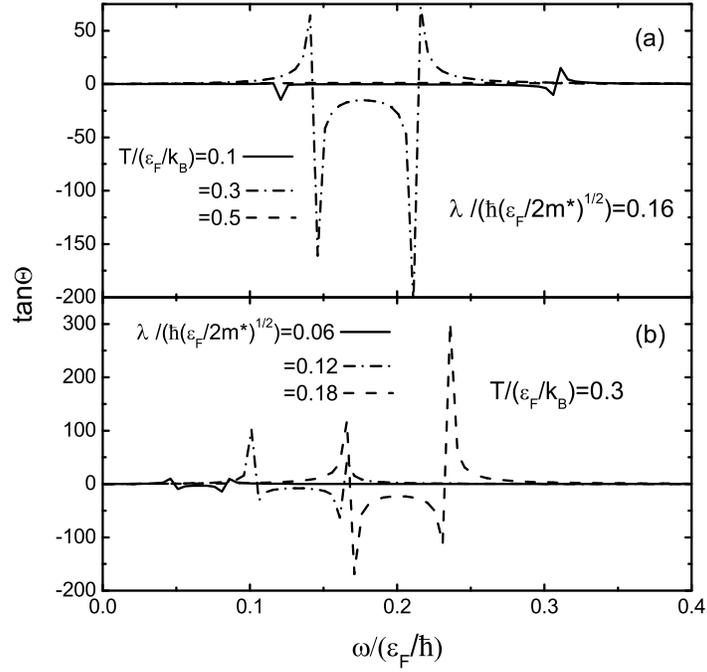} \caption{Dependence with
frequency of the Hall angle. The upper panel (a) displays
variances of the Hall angle for temperatures ranging roughly from
low (solid line) up to room temperature (dash line), for a fixed
Rashba coupling. The lower panel (b) shows variances of the Hall
angle with three different strengths of SO coupling.} \label{fig3}
\end{figure}

\end{document}